\documentclass[10pt,oneside]{article}
\usepackage[english]{babel}
\usepackage{geometry} 
\usepackage[dvipsnames]{xcolor}
\usepackage{color}
\usepackage{graphics}
\usepackage{amsmath}
\usepackage{amssymb}
\usepackage{graphicx}
\usepackage{morefloats}
\usepackage{blkarray}
\usepackage{verbatim}
\usepackage{hyperref}

\title{Detecting early signs of the 2007-2008 crisis in the world trade}
\author{Fabio Saracco$^{1,3}$, Riccardo Di Clemente$^{2,3}$\thanks{\mbox{Corresponding author. \emph{E-mail~address}:~rdicle@mit.edu}}, Andrea Gabrielli$^{1,3}$ \& Tiziano Squartini$^{1,3}$ \\
\footnotesize
$^{1}$\textit{IMT Institute for Advanced Studies Lucca, Piazza S. Ponziano, 6 55100, Lucca, Italy}\\
\footnotesize
$^{2}$\textit{Massachusetts Institute of Technology, Department of Civil and Environmental Engineering, Cambridge, 02139, MA USA}\\
\footnotesize
$^{3}$\textit{Istituto dei Sistemi Complessi-CNR, Via dei Taurini 19, 00185 Rome, Italy}\\
}

\begin{document}
%\date{Published 25-July-2016 on Scientific Reports 6, 532  \href{http://dx.doi.org/10.1038/srep00532}{DOI: 10.1038/srep00532} (2016)}
\maketitle

\begin{abstract}
Since 2007, several contributions have tried to identify early-warning signals of the financial crisis. However, the vast majority of analyses has focused on financial systems and little theoretical work has been done on the economic counterpart. In the present paper we fill this gap and employ the theoretical tools of network theory to shed light on the response of world trade to the financial crisis of 2007 and the economic recession of 2008-2009. We have explored the evolution of the bipartite World Trade Web (WTW) across the years 1995-2010, monitoring the behavior of the system both \emph{before} and \emph{after} 2007. Our analysis shows early structural changes in the WTW topology: since 2003, the WTW becomes increasingly compatible with the picture of a network where correlations between countries and products are progressively lost. Moreover, the WTW structural modification can be considered as concluded in 2010, after a seemingly stationary phase of three years. We have also refined our analysis by considering specific subsets of countries and products: the most statistically significant early-warning signals are provided by the most volatile macrosectors, especially when measured on developing countries, suggesting the emerging economies as being the most sensitive ones to the global economic cycles.

\end{abstract}

\section*{Introduction\label{introduction}}

During 2007 one of the most striking financial crises of the century has manifested itself. This major event has motivated researchers and institutions to devote a large effort for better understanding economic and financial systems evolution, with the aim of detecting reliable signals on upcoming critical events enough in advance to allow regulators to plan effective policies to contrast and overcome them \cite{hatz,silvio,tarik,myearly,guilera,forum}. The vast majority of analyses, however, has focused on financial systems \cite{hatz,silvio,tarik,myearly,stani} and little theoretical work has been done, so far, on the economic counterpart \cite{guilera}, though the definition of better early-warning indicators for the economic systems is advocated by many organizations, as the International Monetary Fund \cite{nber,nber2,nber3}, the United Nations \cite{shelby,un} and the national central banks \cite{czech}. In particular, little effort has been done to go beyond a detailed description of the effects of the financial crisis on the world trade (and of the subsequent economic recession), after the triennium 2007-2009 \cite{shelby,forum,ricca,nber}.

\color{black}With the aim of filling this gap, and complementing the existing vast amount of literature on financial markets, in the present paper we analyse the bipartite World Trade Web (hereafter WTW)\cite{Hidalgo2007, Hidalgo2009, Tacchella2012, Cristelli2013}, by employing a novel method to assess the statistical significance of a number of topological network properties across the period 1995-2010 \cite{mybipmethod}: more specifically, we focus on a recently-proposed class of bipartite motifs, studying their evolution across the aforementioned period.

Our temporal window of sixteen years allows us to monitor the system both \emph{before} the year 2007 (i.e. during the years preceeding the crisis) and \emph{after} it (i.e. during the critical and post-critical phases): as we will show, the considered family of motifs clarifies the role played by the year 2007 in the economic framework of the world trade. Our analysis suggests that this year marks a \emph{crossover} from a phase characterized by a steep increase of randomness of the WTW topology - and individuates 2007 as the \emph{last} year of this critical phase started \emph{earlier} in time - to a phase during which a stationary regime seems to be finally reached. Indeed, the abundances of the considered family of motifs point out that the crisis explicitly manifests itself after a period of four years during which the WTW has undergone a dramatic structural change.

Notably, the class of motifs introduced in \cite{mybipmethod} allows us to focus on \emph{nodes subsets}, too. In particular, we analyse the evolution, across the same temporal window, of subgroups of nodes predicated to show remarkable economic homogeneities, with the aim of detecting the effects of the worldwide crisis on various segments of trade - that we consistently define by adopting a classification into macrosectors, as agriculture, manufacture, services, etc. - and on the exports of several national economies \cite{ricca} - grouped into BRICS, PIGS, G7, etc. Our analysis evidences that some sectors/groups of countries are more sensitive to the cycles of the worldwide economy, providing robust early-warning signals of the 2007 crisis (and confirming the trends individuated at the global level); others, instead, provide little information on its build-up phase. Surprisingly, our study reveals also the existence of subsets of nodes which do not show any relevant internal correlation throughout the whole 1995-2010 window, thus questioning the correctedness of the reasoning leading to the individuation of such groups as homogeneous sets.

The rest of the paper is organized as follows. Section Data describes the data set used for the present analysis; section Methods sums up the algorithm implemented for the present analysis, illustrating the null model and the statistical indicators we have considered; section Results presents the results of our analysis that we comment in the section Conclusions.

\section*{Results\label{results}}

\subsection*{Data\label{data}}

In this paper we represent the WTW as a bipartite, undirected, binary network where countries and products constitute the nodes of the two different layers, obeying the restriction that links connecting nodes of the same layer are not allowed.

The data set used for the present analysis is the BACI World Trade Database \cite{gaul}. The original data set contains information on the trade of 200 different countries for 5000 different products, categorised according to the 6-digits code of the Harmonized System 2007 \cite{database}. The products' division in sectors, instead, follows the UN categorisation \cite{onu}; in order to apply it, we have converted the HS2007 code into the ISIC revision 2 code at 2-digits \cite{conversion}. In what follows we have proxied countries' trade volumes by considering exports (measured in US dollars): after the cleaning procedure operated by COMTRADE \cite{gaul} and the application of the RCA threshold (Revealed Comparative Advantage - see the Appendix for further details), we end up with a rectangular binary matrix $\mathbf{M}$ (i.e. the biadjacency matrix of our bipartite, undirected, binary network) whose dimensions \color{black} are $C$ and $P$, respectively indicating the total number of countries ($C\simeq 140$ for all years of our temporal window) and the total number of products ($P=1131$) considered for the present analysis. The matrix generic entry $m_{cp}$ is 1 if country $c$ exports an amount of product $p$ above the RCA threshold; otherwise, $m_{cp}=0$. In economic terms, each row represents the basket of exported products of a given country; analogously, each column represents the set of exporters of a given product.

\subsection*{Closed motifs analysis}\label{sec:cma}

\begin{figure}[t!]
\begin{center}
\includegraphics[width=0.5\textwidth]{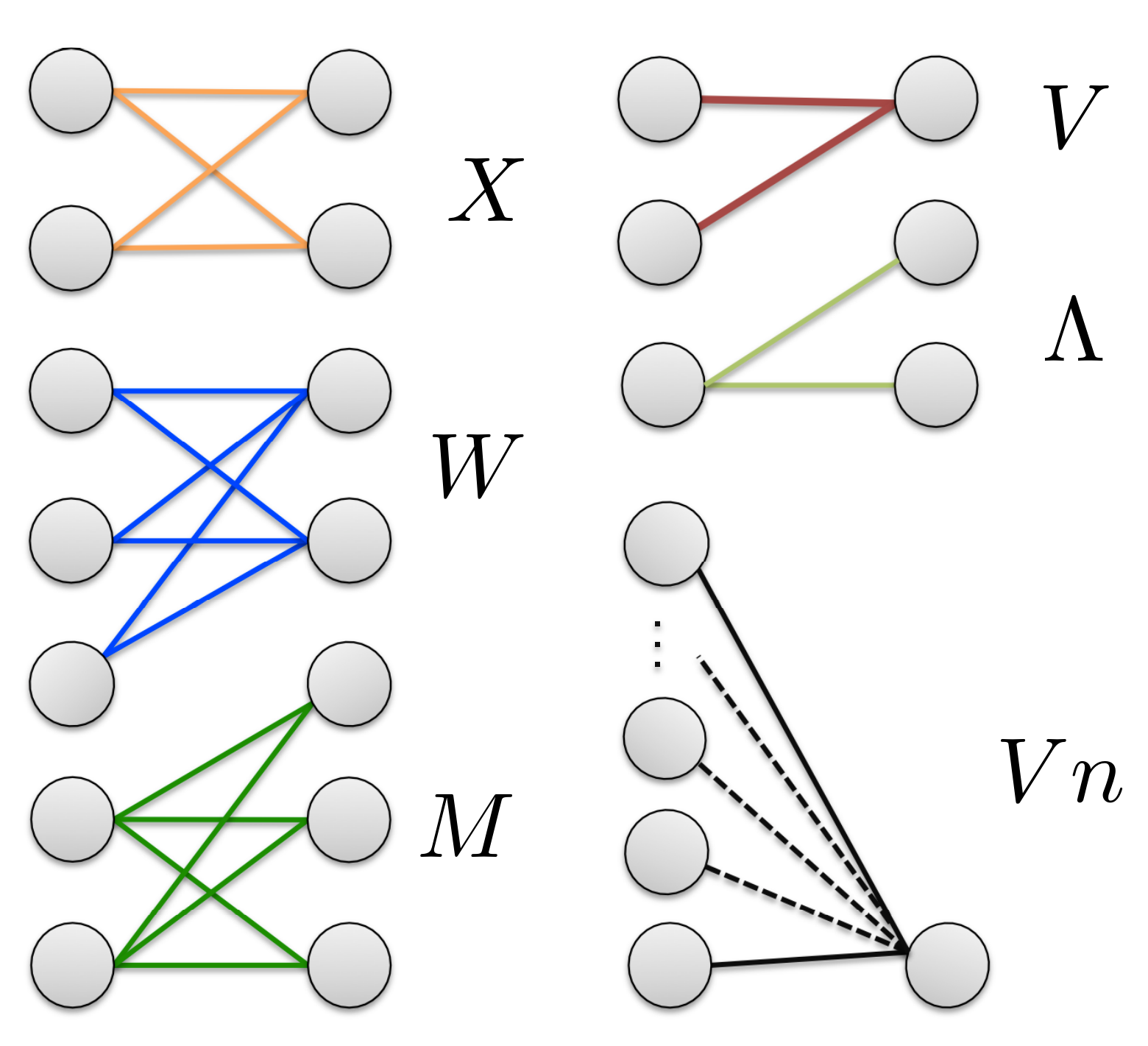}
\caption{Examples of motifs for bipartite networks. Countries are reported in the left layer, products in the right layer. The V$n$ motifs quantify the degree of similarity of the exports of $n$ arbitrarily chosen countries, generalizing the measure provided by V-motifs (applicable to pairs of countries only).}
\label{motifsapp}
\end{center}
\end{figure}

The quantities we have considered for the present analysis are the bipartite motifs introduced in \cite{mybipmethod}, i.e. the V-, $\Lambda$-, X-, W- and M-motifs (see fig. \ref{motifsapp}). While V-motifs account for the total number of pairs of countries exporting the same products, i.e. $N_V(\mathbf{M})=\sum_{c<c'}V_{cc'}$ with $V_{cc'}=\sum_pm_{cp}m_{c'p}$, $\Lambda$-motifs account for the total number of pairs of products in the basket of the same countries, i.e. $N_\Lambda(\mathbf{M})=\sum_{p<p'}\Lambda_{pp'}$ with $\Lambda_{pp'}=\sum_cm_{cp}m_{cp'}$. However, as shown in the Methods section, V- and $\Lambda$-motifs provide a rather limited information on the structure of the bipartite WTW: for this reason, we have focused our analysis on more complex (closed) patterns, i.e. the X-, W- and M-motifs, which can be compactly expressed as combinations of V-, or $\Lambda$-, motifs. In economic terms, our motifs provide a measure of the overlap of countries' baskets of products.

In particular, X-motifs can be defined as combinations of pairs of V-motifs (or $\Lambda$-motifs), subtending the same two countries and products (see fig. \ref{motifsapp}):
\begin{equation} 
N_{X}(\mathbf{M})=\sum_{c<c'}\binom{V_{cc'}}{2}=\sum_{p<p'}\binom{\Lambda_{pp'}}{2}.
\label{xx}
\end{equation}

\indent In other words, X-motifs quantify the co-occurrence of any two countries as producers of the same couple of products (and, viceversa, the co-occurrence of any two products in the basket of the same couple of countries). Allowing for a higher number of interacting nodes, W-motifs and M-motifs (see fig. \ref{motifsapp}) can be defined, respectively enlarging the set of countries and the set of products to triples: 
\begin{equation}
N_{W}(\mathbf{M})=\sum_{p<p'}\binom{\Lambda_{pp'}}{3},\:N_{M}(\mathbf{M})=\sum_{c<c'}\binom{V_{cc'}}{3}.
\label{wm}
\end{equation}

In order to assess the statistical significance of our observations, the definition of a proper statistical benchmark (or null model) is needed: the aim of null models is precisely that of washing out the contribution of some lower-order constraints to the observed network structure. Here we follow \cite{mybipmethod} (which builds upon the works \cite{mymethod,Newman2003,Newman2004,Fronczak2014} dealing with null models for monopartite networks) and consider the benchmark provided by the {\it bipartite configuration model} (BiCM hereafter). In analogy with the monopartite case, such a null model is defined by the constraints represented by the degree sequence of nodes (i.e. the number of connections for each node) on both layers (see Methods for further analytical details).

\begin{figure*}[t!]
\begin{center}
\includegraphics[width=\textwidth]{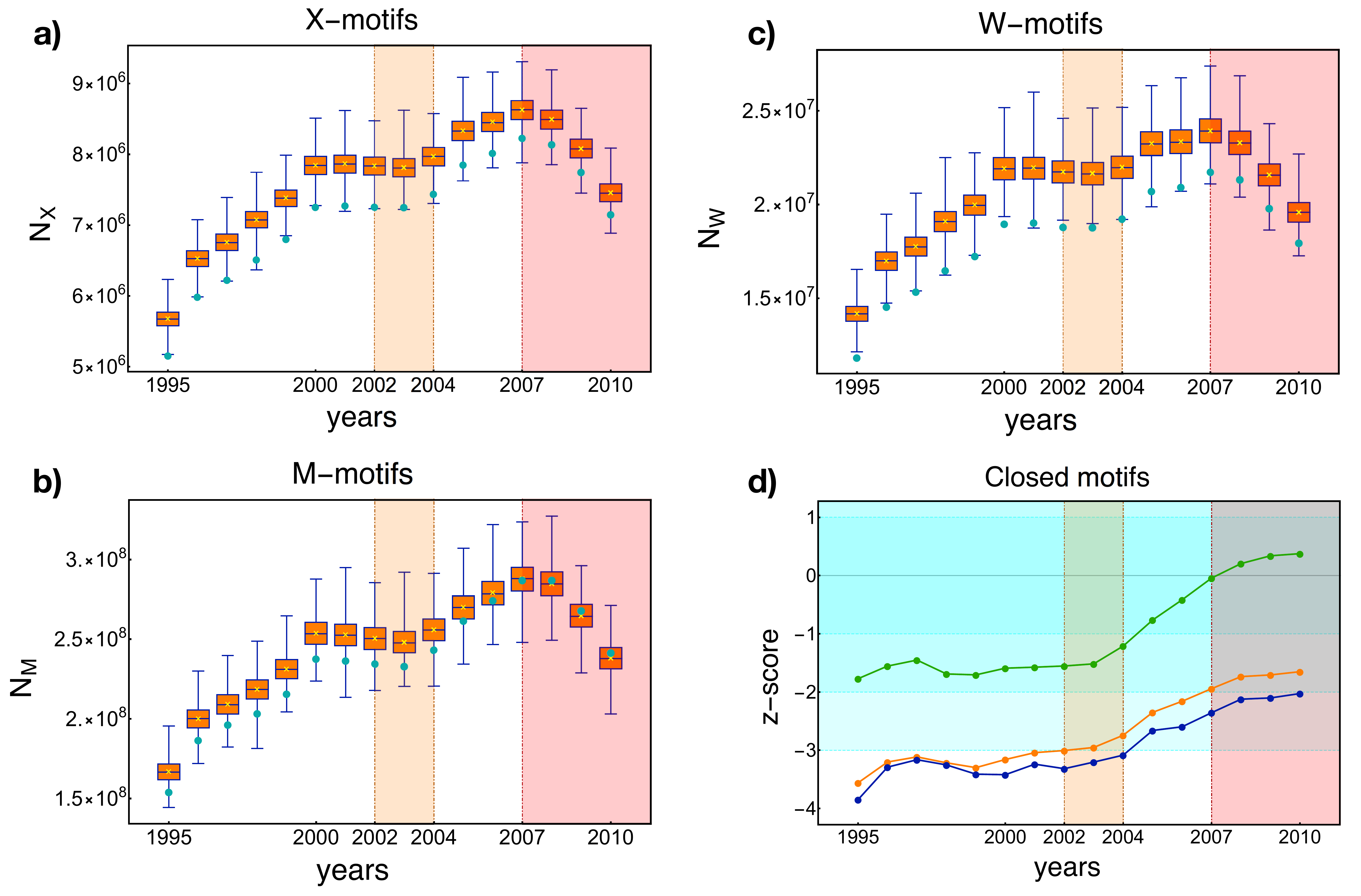}
\caption{Evolution of X-motis, W-motifs and M-motifs across the period 1995-2010. Panels a), b) and c) show the box plots of the X-, M- and W-motifs respectively, across the considered time window. The mean of the year-specific BiCM-induced ensemble distribution is indicated as a yellow cross; light-blue dots represent the observed abundances of motifs. Panel d) shows the evolution of the motifs $z$-scores (X-motifs in orange, W-motifs in blue, M-motifs in green). Beside pointing out the peculiarity of the quinquennium 2003-2007, suggesting it as the most critical phase undergone by the WTW, the $z$-scores of our bipartite motifs make the system evolution after 2007 inspectable as well, indicating the presence of a phase characterized by an attenuation of the crisis effects.}
\label{motifsz}
\end{center}
\end{figure*}

Let us start by comparing the observed abundances of our X-, W- and M- motifs in the real network with the corresponding expected values in the null case plotting, first, the motifs-specific box plots: as specified in the Methods section, the latter are intended to sum up the year-specific BiCM-induced ensemble distributions through a bunch of percentiles (see fig. \ref{motifsz}, where the distribution mean is explicitly shown as a yellow cross, while the observed abundances of motifs are represented by turquoise-colored dots).

The observed trends of all motifs suggest the presence of four distinct temporal phases (1995-2000, 2000-2003, 2003-2007 and 2007-2010), the last one starting with a sudden trend inversion in correspondence of 2007: this evidences that the evolution of our motifs correctly points out the critical character of the year 2007, after which an overall contraction of the worldwide trade (which continues at a higher rate in the following years) is clearly visible. Such a contraction is related to a decrease of the number of trades among countries, since the evolution of the total number of links is characterized by a similar trend: this finding highlights that the topological structure of the WTW has indeed started changing since 2007, in turn confirming that the financial crisis has also affected the world economy \cite{nber,nber2,nber3,rein,lev}.

%\begin{figure*}[t!]
%\begin{center}
%\includegraphics[width=\textwidth]{z_motifs.pdf}
%\caption{Evolution of X-motis, W-motifs and M-motifs across the period 1995-2010. Panels a), b) and c) show the box plots of the X-, M- and W-motifs respectively, across the considered time window. The mean of the year-specific BiCM-induced ensemble distribution is indicated as a yellow cross; light-blue dots represent the observed abundances of motifs. Panel d) shows the evolution of the motifs $z$-scores (X-motifs in orange, W-motifs in blue, M-motifs in green). Beside pointing out the peculiarity of the quinquennium 2003-2007, suggesting it as the most critical phase undergone by the WTW, the $z$-scores of our bipartite motifs make the system evolution after 2007 inspectable as well, indicating the presence of a phase characterized by an attenuation of the crisis effects.}
%\label{motifsz}
%\end{center}
%\end{figure*}

Turquoise-colored dots in panels a, b, c of fig. \ref{motifsz} represent the abundances of closed motifs measured on the real network. Clearly, the results of a purely empirical analysis can point out the peculiarity of the year 2007 only \emph{a posteriori}, i.e. only \emph{after} that the economic recession has manifested itself: stated otherwise, no early-warning signals of the upcoming crisis are provided by the observed trends of motifs. Rather, a na\"ive estimation of the future trends of such indicators, solely based on their pre-2007 history, would have probably predicted a trend increase, thus failing in detecting the approaching crisis (similarly, many economic indicators registered a growth in virtually all countries during the years preceeding the crisis \cite{nber}). 

A deeper insight on the system evolution can be obtained by quantifying the discrepancy between the observed abundances of motifs and the corresponding predicted values under the aforementioned null model, i.e. the BiCM (see panels a, b, c of fig. \ref{motifsz}). Indeed, by considering the variation of the turquoise dot position within the interquantile interval, one realizes that the statistical significance of the observed abundances progressively diminishes as 2007 approaches. A compact way of quantifying this tendency is represented by the computation of the motifs $z$-scores, as panel d of fig. \ref{motifsz} shows (the ensemble distributions of our motifs abundances are very accurately approximated by Gaussians - see fig. \ref{check} in Appendix - allowing for the usual interpretation of $z$-scores - see also Methods). The analysis of $z$-scores reveals us which one of the aforementioned four phases are genuinely statistically significant, drawing an interesting picture. Worth to be noticed is the rate of increase of randomness across 2003, which seems to play a discriminating role between two distinct phases (thus ruling out any intermediate trend inversions): from 1995 to 2003, the WTW is described by the practically constant level of statistical significance of all the considered motifs; from 2003 on, the WTW undergoes an evident structural change.

In particular, the WTW becomes more and more compatible with the BiCM during the quinquennium 2003-2007: the transition towards less negative $z$-scores indicates a progressive increase of the abundance of X-, W- and M- motifs during the considered time-period, pointing out the tendency to establish a number of motifs more and more similar to the expected one. Since our null model preserves only the degree sequence while randomizing the rest of the network topology, this implies that 2003-2007 can be regarded as a period characterized by a steep increase of randomness (and, as a consequence, by a loss of internal structure), leading e.g. the observed abundance of M-motifs to be exactly reproduced by our null model in 2007. The evolution of our motifs $z$-scores contributes to shed light on the actual role played by the year 2007 in the economic context of world trade. As our analysis suggests, the truly crucial phase experienced by the WTW is represented by the quinquennium 2003-2007: the year 2007, as fig. \ref{motifsz} illustrates, is the ending point of such a phase started \emph{earlier} in time. 

The loss of internal structure affecting the international trade is related to an increasing similarity of the export basket of different countries \cite{patt}. Since this effect is not imputable to an increase of the number of country-specific exports before 2007 - our null model automatically discounts it - it seems to represent a genuine tendency of countries, in particular of emerging economies, which tend to align their exports with those of the advanced ones, as it has been recently pointed out \cite{patt}. This tendency may have made the whole system more fragile in the years preceding the worldwide crisis, thus negatively impacting on the systemic resilience to propagation of financial and economic shocks \cite{resil}. Interestingly, it has been recently noticed that an improvement of the world-economy stability may be gained upon adopting strategies similar to those of ecosystems, which call for a larger (economic) adaptability based on production and export variety \cite{resil}.

\subsection*{Sectorial analysis}

In the previous subsection we have considered the correlations between nodes only at a global level; let us now focus on specific groups of products or countries. Notably, all the aforementioned motifs can be defined onto specific subgroups of nodes as well. For instance, the correlations between products within the same sector $s$ can be evaluated by computing $N_{\Lambda^{s}}=\sum_c\sum_{\substack{p<p'\\p,p'\in s}}m_{cp}m_{cp'}$. Similarly, the abundance of restriced X-, W- and M-motifs can be computed by applying definitions \ref{xx} and \ref{wm} to restricted V- and $\Lambda$-motifs. We would like to stress that, whenever the computation of V- and $\Lambda$-motifs is restricted to subsets of nodes, their z-scores are no longer functions of global quantities (as the total number of links, the total number of nodes, etc. - see the Methods section), thus providing non-trivial information on the network structural organization.

\subsubsection*{Worldwide macrosectors}

Let us consider the subsets of products known as ``macrosectors'' in the economic jargon (see the Appendix for the list of sectors considered in the present paper). According to the economic theory, macrosectors are closely related to the so-called vertical markets, i.e. markets in which similar products are developed using similar methods. From the perspective of trade, this leads to the assumption that products belonging to the same sector are (produced and) exported together; this, in turn, translates into the expectation of detecting macrosector-specific patterns showing statistically significant evidences of the predicated similarity and, by converse, leads to interpret systematic changes of such a significance as sectorial early-warning signals.

\begin{figure}[t!]
\begin{center}
\includegraphics[width=0.45\textwidth]{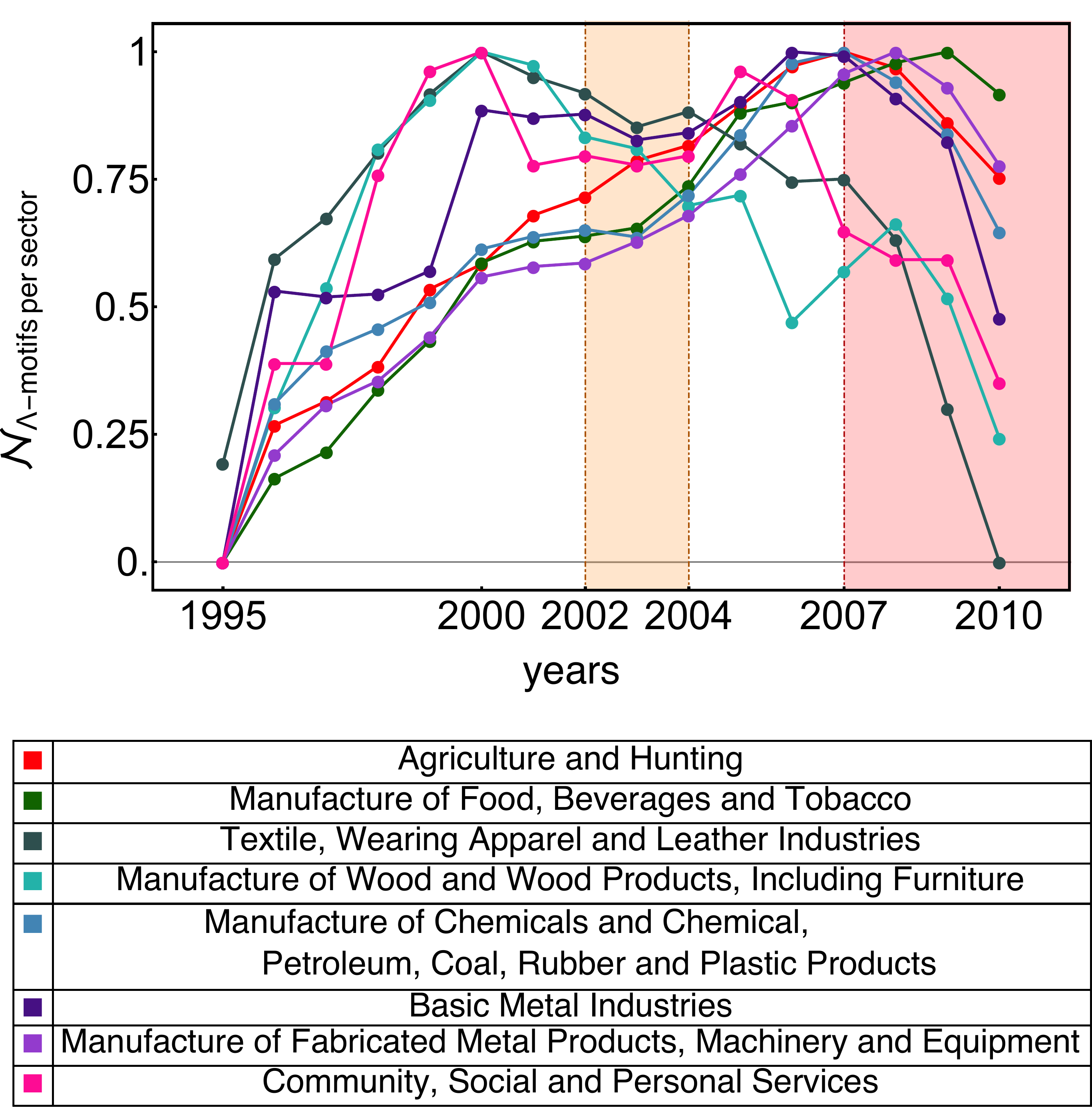}
\caption{Evolution of the observed abundances of restricted $\Lambda$-motifs, measured within the products belonging to the same sector (see the legenda in the bottom panel). The trend inversions in correspondence of the year 2007 indicate a contraction of trade in, practically, all its segments. Values for each specific year, $y$, have been normalized to range within $[0,1]$ according to the prescription $\mathcal{N}_\Lambda^{y}=(N_\Lambda^{y}-\min\{\Lambda\}_{95}^{10})/(\max\{\Lambda\}_{95}^{10}-\min\{\Lambda\}_{95}^{10})$.}
\label{lambda_tot}
\end{center}
\end{figure}

\color{black}According to many observers \cite{nber,nber2,nber3,czech}, virtually all sectors and countries have been exposed to the effects of the worldwide crisis of 2007. Fig. \ref{lambda_tot} shows the evolution of the observed abundances of restricted $\Lambda$-motifs (measured within the products belonging to the same sector) for all macrosectors: our plots confirm the tendency highlighted in \cite{nber,nber2,nber3,czech}, showing trend inversions in correspondence of the year 2007 which indicate a contraction of trade in, practically, all its segments. Nonetheless, only part of the latter seems to provide early-warning signals of the 2007 crisis.

\begin{figure*}[t!]
\begin{center}
\includegraphics[width=\textwidth]{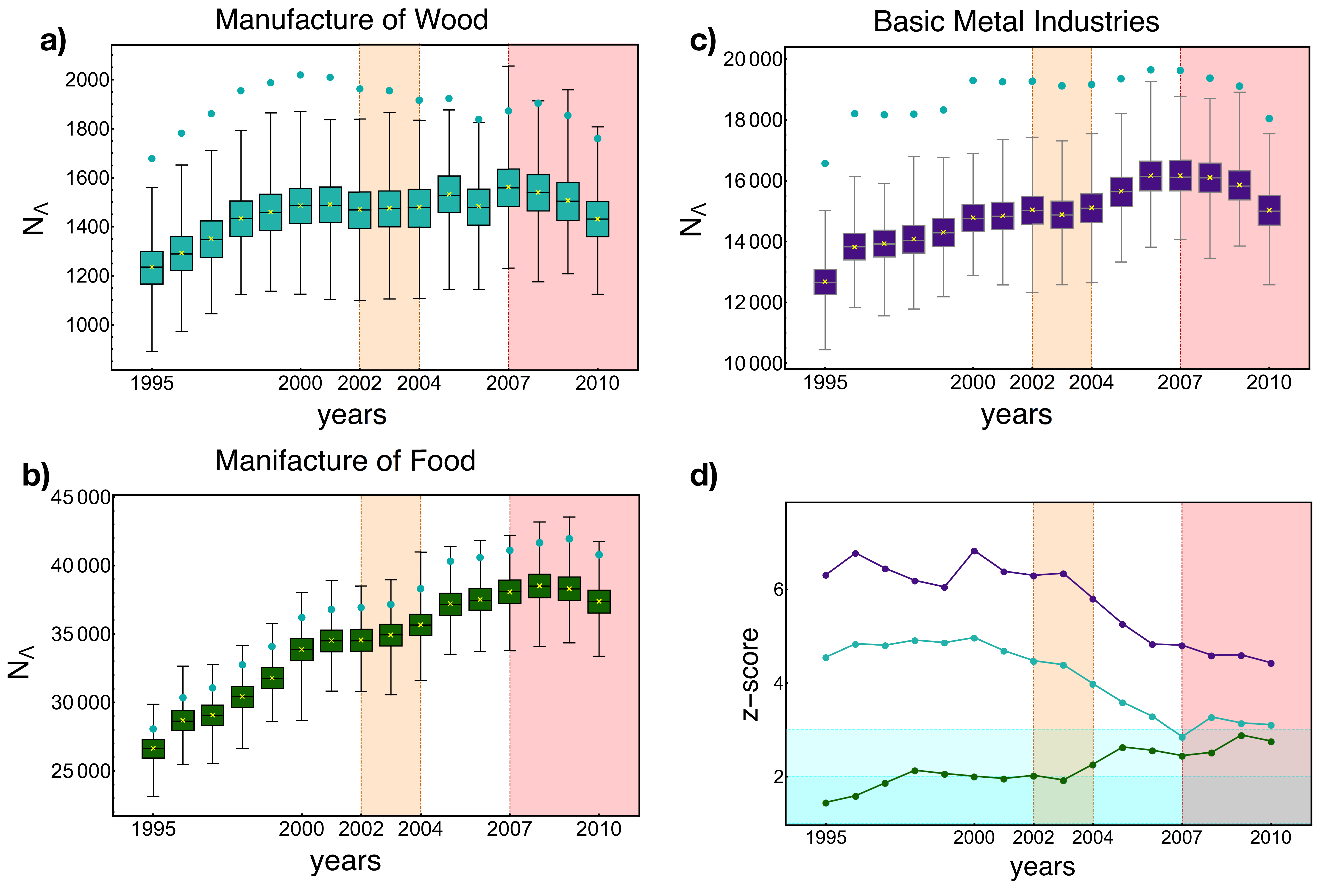}
\caption{Evolution of restricted $\Lambda$-motis across the period 1995-2010. Panels a), b) and c) show the box plots of the $\Lambda$-motifs of sectors \textcolor{cyan}{$\bullet$} - manufacture of wood products, \textcolor{RoyalPurple}{$\bullet$} - basic metal industries and \textcolor{OliveGreen}{$\bullet$} - manufacture of food, beverages and tobacco across the considered time window. The mean of the year-specific BiCM-induced ensemble distribution is indicated as a yellow cross; light-blue dots represent the observed abundances of motifs. Panel d) shows the evolution of the motifs $z$-scores. Even at the macrosectors level, the peculiarity of the quinquennium 2003-2007 is apparent. Interestingly, the most robust early signals of the WTW structural change are detectable upon inspecting the most volatile economic sectors.}
\label{productsz}
\end{center}
\end{figure*}

The most evident signals are provided by sectors ``manufacture of wood products'' and ``basic metal industries'' (see fig.\ref{productsz}). From a purely economic point of view, it has been noticed that the metal market has been among the most prone ones to prices fluctuations across 2008, representing one of the sectors characterised by the largest fall of prices \cite{nber,nber2,nber3,french}; this has been interpreted as the consequence of the deceleration in industrial production, which has negatively affected the metal demand \cite{nber,nber2,nber3,french}. This seems to be mirrored by the evolution of our $z$-score, which points out the progressively less significant tendency to trade basic metals from 2003 on. The manufacturing sector seems to have been characterized by the same fate, witnessing a significant decrease in its trade as well; this has been related to the involvement of developing countries into the global crisis, i.e. the most active ones in the manufacturing sector of raw materials (especially the manufacture of wood) because of the ongoing industrialization process \cite{nber,nber2,nber3,wood}. Remarkably, both sectors show evidence of the post-2007 randomness stabilization, thus confirming their key role as very sensitive trade-segments to the cycles of the worldwide economic activity \cite{nber,nber2,nber3}.

Particularly interesting is the behavior of the sector ``manufacture of food, beverages and tobacco'' whose $z$-score trend keeps increasing throughout our temporal window (see panel d of fig.\ref{productsz}). In particular, it jumps to levels of progressively higher statistical significance: one of these jumps takes place precisely in 2003, thus confirming the peculiar character of this year. This finding seems to be mirrored by the growing demand of food which has been observed in the years before the crisis and which has been related to the prices fluctuations of other trade-segments \cite{food,food2}.

The sectors we have considered in this section are defined by products characterized by a moderately high level of internal similarity: nonetheless, all of them provide early indicators of the worldwide crisis of 2007, pointing out the peculiarity of the preceeding quinquennium. It is interesting to compare the behavior of such sectors with those characterized by the most positive values of the corresponding $z$-scores (i.e. ``manufacture of chemicals'', ``manufacture of fabricated metal products'' and ``textile''). As evident upon inspecting fig. \ref{productsz3} in the Appendix, the latter seem to be insensitive to the approaching crisis, showing remarkably stable trends (increasing, decreasing or flat) throughout our whole temporal window.

\begin{figure*}[t!]
\begin{center}
\includegraphics[width=\textwidth]{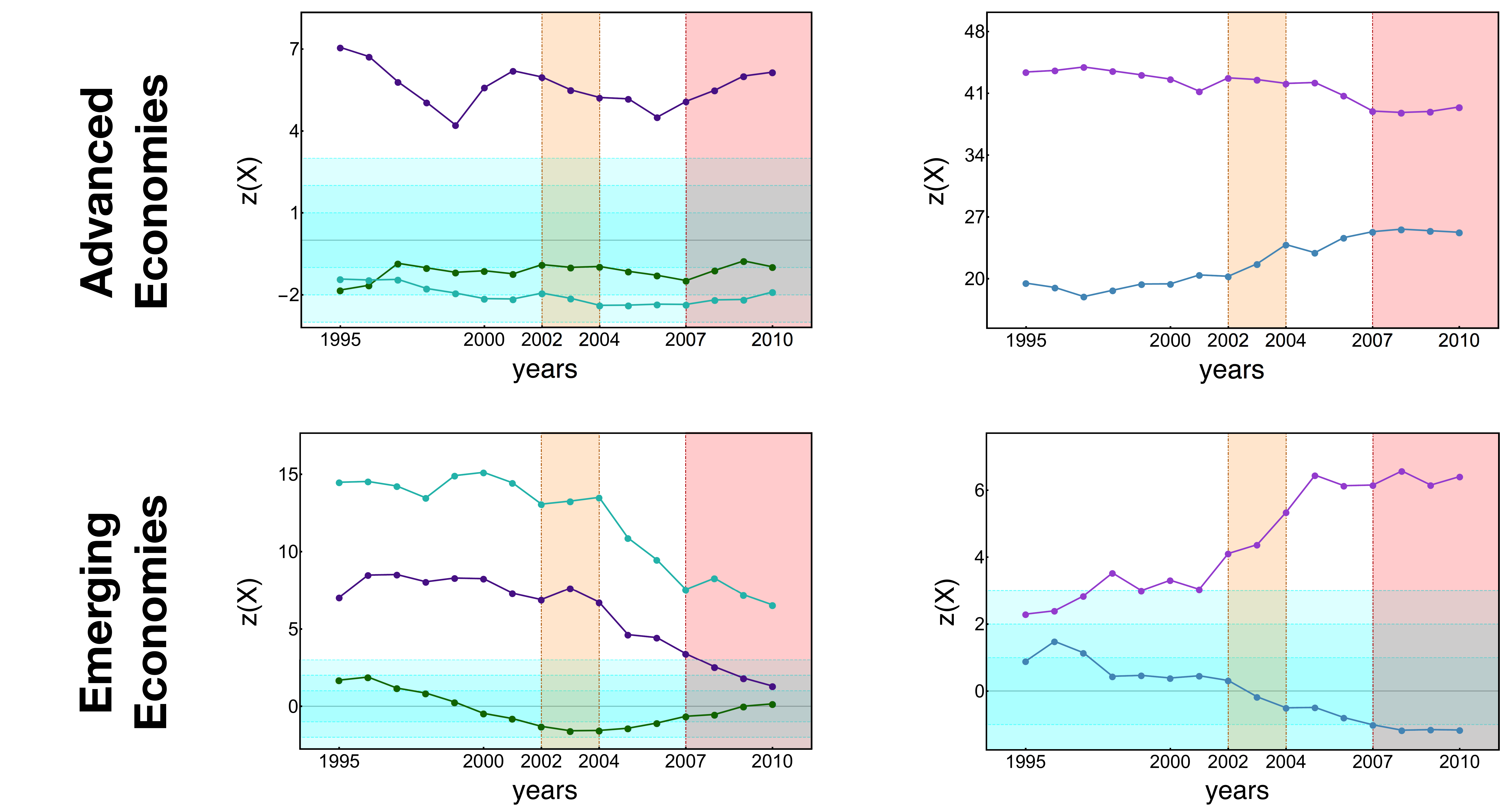}
\caption{Evolution of $z$-scores of restricted X-motifs across the period 1995-2010, measured on sectors \textcolor{cyan}{$\bullet$} - manufacture of wood products, \textcolor{RoyalPurple}{$\bullet$} - basic metal industries, \textcolor{OliveGreen}{$\bullet$} - manufacture of food, beverages and tobacco, \textcolor{Orchid}{$\bullet$} - manufacture of chemicals, \textcolor{Gray}{$\bullet$} - manufacture of fabricated metal products, for advanced economies (top panels) and emerging economies (bottom panels). Interestingly, the most robust early-warning signals of the WTW structural change are provided by emerging economies, for all the considered sectors.}
\label{productsz2}
\end{center}
\end{figure*}

\subsubsection*{Advanced VS developing economies}

So far, we have considered sectors only at the worldwide level. However, the large heterogeneity of the countries composing our data set pushes us to carry on a more detailed analysis. In this subsection we have focused our attention on one third of the total number of nations, considering two distinct groups of world countries \cite{nber}: the advanced economies (Australia, Canada, Czech Republic, Denmark, Euro area, Hong Kong, Israel, Japan, Korea, New Zealand, Norway, Singapore, Sweden, Switzerland, Taiwan, United Kingdom, United States) and the developing economies (Argentina, Brazil, Bulgaria, Chile, China, Colombia, Estonia, Hungary, India, Indonesia, Latvia, Lithuania, Malaysia, Mexico, Pakistan, Peru, Philippines, Poland, Romania, Russia, Slovak Republic, South Africa, Thailand, Turkey, Ukraine, Venezuela).

Repeating the sectorial analysis for these two groups of countries (through the restricted X-motifs) leads to the results reported in fig. \ref{productsz2}. What emerges is that, while the advanced economies show rather stable trends, the most evident early-warning signals are provided by the developing economies, for all the considered sectors: in fact, they show the most evident losses of statistical significance (e.g. from $z\simeq14$ to $z\simeq6$ for the ``manufacture of wood products''), for all the considered sectors since 2003. Remarkably, even the sectors which at the worldwide level do not evidence any structural changes (see fig. \ref{productsz3} in the Appendix), when decoupled into advanced and emerging economies, corroborate the aforementioned finding.

From the economic point of view, this result has two important consequences. First, it indicates that the 2007 crisis has deeply impacted on developing economies and, statistically speaking, in a much more pronounced way than on advanced economies, despite the previsions on the relative safety of emerging markets \cite{nber,nber2,nber3}. Second, emerging economies seem to be much more sensitive to the economic cycles than advanced economies, thus representing useful indicators of the health conditions of the worldwide trade \cite{emerge,emerge2}.

\subsection*{Countries-specific analysis}

Let us now move to considering groups of countries, whose correlations have been measured through the restricted V-family of motifs. More explicitly, the degree of similarity of exports of, e.g., countries $i$, $j$ and $k$ can be quantified by calculating $N_{V^{ijk}}=\sum_pm_{ip}m_{jp}m_{kp}$ and analogously for more numerous groups.

As will clearly appear in what follows, the detection of early-warning signals when considering groups of national economies is less straightforward; nonetheless, our approach can be useful in order to quantify the economic similarities (in terms of exported products) of given groups of countries during our temporal window, especially those which have been traditionally regarded as sharing common strategies (e.g. political and administrative ones) for developing their economic growth \cite{ricca2,ricca3}.

\begin{figure*}[t!]
\begin{center}
\includegraphics[width=\textwidth]{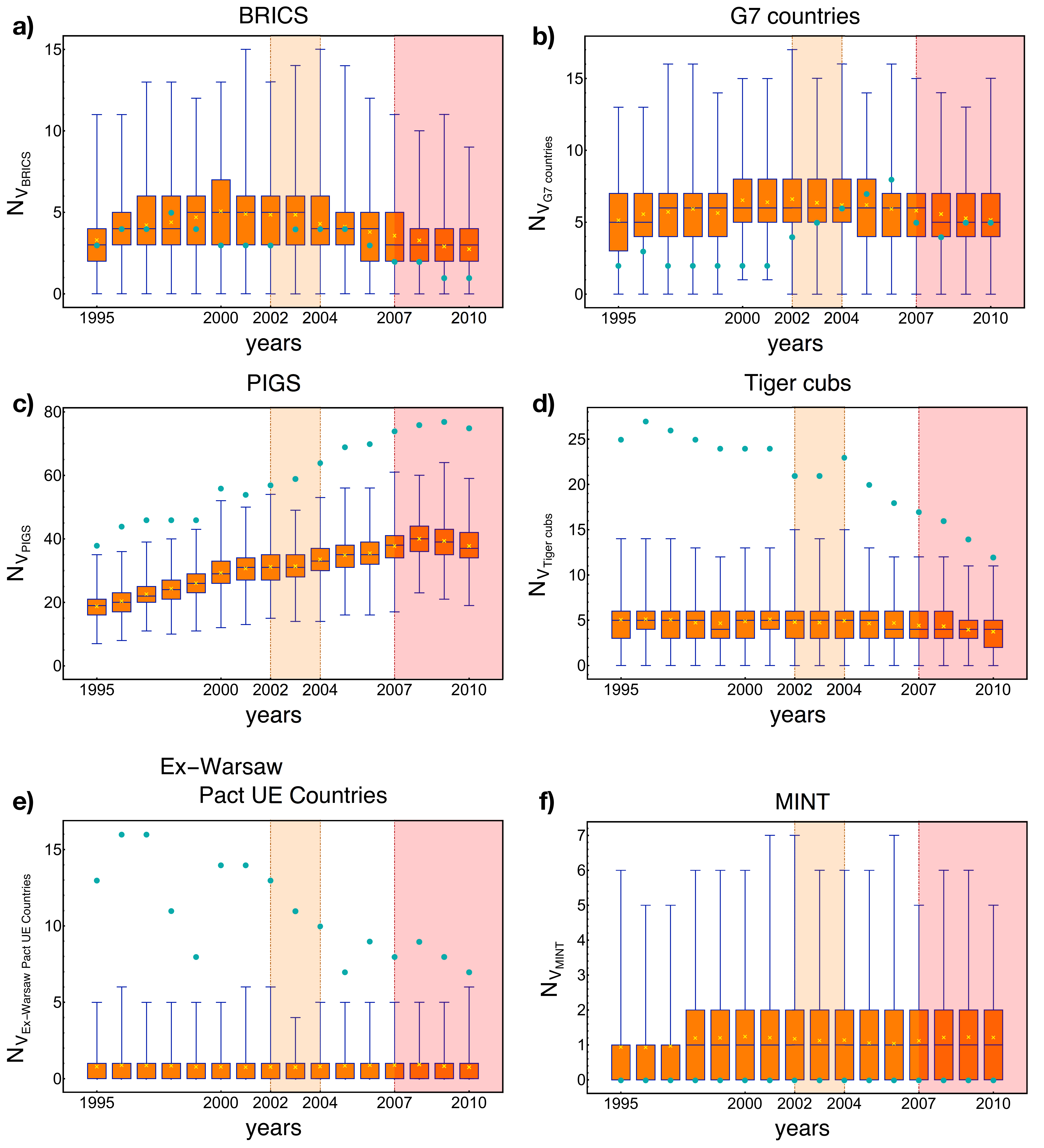}
\caption{Evolution of V-motifs defined for six specific subgroups of countries (BRICS countries, G7 countries, PIGS countries, Tiger Cubs, Ex-Warsaw Pact UE countries and MINT countries). Box plots sum up the year-specific BiCM-induced ensemble distribution, whose mean is indicated as a yellow cross; light-blue dots represent the observed abundances of motifs. Different groups of countries show remarkably different levels of similarity: while some groups are characterized by motifs which are always consistent with our null model, thus questioning the economic relevance of such sets (as BRICS countries), other groups are steadily significant (as PIGS countries).}
\label{ecogroups}
\end{center}
\end{figure*}

Since not all the considered groups of countries are characterized by normal ensemble distributions, a countries-specific analysis based on $z$-scores is not feasible. However, upon inspecting the evolution of the relative positions of the observed and expected abundances of V-motifs within the chosen groups, some clear tendencies appear. Fig. \ref{ecogroups} shows the result of our motifs analysis for six groups of countries: remarkably, different groups of countries are characterized by strongly different statistical evidences of similarity.

BRICS countries (Brazil, Russia, India, China and South Africa), for example, are characterized by motifs which are always consistent with our null model; this means that, at least in the considered period, the similarity of BRICS countries is completely encoded into their degree sequences, suggesting no further peculiarities of their export baskets. The lack of statistical significance confirms what already found in e.g. \cite{Tacchella2012,Cristelli2013} by running a completely different analysis: BRICS countries were found to follow different evolutions (sometimes even divergent), thus questioning the relevance of such a group, predicated to represent similar emerging economies. Moreover, in the last two years of our data set, the observed abundance of BRICS motifs is found to be in the left tail of the distribution, indicating that BRICS countries tend to be characterized by weaker correlations than expected, thus seemingly confirming a tendency to adopt different export strategies.

G7 countries (United States, Japan, Germany, France, United Kingdom, Italy, Canada), on the other hand, are found in the left tail of our distributions during the first years of our data set, close to the significance border of the 0.15th percentile. This means that \emph{before} 2002 the differences concerning the exports of G7 countries were somehow dominant; they have become steadily less significantly different from 2003 to 2007 and fully consistent with our null model in the last years of our data set (confirming the increasing level of randomness detected at the global level).

PIGS countries (Portugal, Italy, Greece, Spain), instead, represent a significant grouping for all years. Contrarily to BRICS, their similarity increases during the quinquennium 2003-2007, not showing any decrease after 2007. This suggests that the exports of such countries have been affected by the same tendency that has been detected at a global level. Interestingly, it has been noticed that PIGS were hit the hardest in Europe \cite{pigs}: this seems to confirm the role that the less solid national economies (in this case within the Euro area) may play as sensitive indicators of crises.

Interestingly, Tiger Cubs (Malaysia, Indonesia, Philippines, Thailand) show the same tendency of the european countries formerly within the Warsaw Pact (Bulgaria, Czech Republic, Estonia, Hungary, Latvia, Lithuania, Poland, Slovakia) to lose their peculiarities. For the first group of countries this may reflect the fact that they have undergone a remarkably similar industrialization process in the first years of our data set, which has then led to a progressive diversification with the passing of time. For the second group of countries, on the other hand, such a tendency is explainable upon considering the presence of a similar national economic organization, whose effects have persisted well after the fall of the Berlin Wall; only the last quinquennium seems to witness a genuine diversification of the exports of the aforementioned eastern-Europe countries. Intriguingly, the results of the analysis presented in \cite{scienceof} confirm this tendency, even at the level of the scientific production.

Our approach can be also used to monitor other groups of countries, believed to constitute genuinely similar economic realities in the near future. An example is provided by the so-called MINT countries (Mexico, Indonesia, Nigeria, Turkey) \cite{O'Neill}, representing a small portion of the larger Next 11 group (Bangladesh, Egypt, Indonesia, Iran, Mexico, Nigeria, Pakistan, Philippines, Turkey, South Korea, Vietnam). As shown in fig. \ref{ecogroups}, no appreciable signal is detected within the considered temporal interval: in other words, our data set does not provide any evidence in favor of such a grouping, even more clearly than for the other aforementioned ones.

\section*{Discussion}

Our analysis sheds light on the evolution of the bipartite World Trade Web\cite{Hidalgo2007, Hidalgo2009, Tacchella2012, Cristelli2013} across 2007, through the comparison of the observed trends of a recently-defined family of bipartite motifs with their expected values, under an appropriate null model: following the approach in \cite{mybipmethod}, we constrain the degree of each node and assess the statistical significance of such higher order topological features. Thus, detecting statistically-significant {\it discrepancies} between observed and expected values is interpreted as the signature of structural changes, not encoded into the variation of the degree sequence. Upon analyzing the temporal evolution of these discrepancies for our family of motifs (see also the Methods for results on additional indicators such as nestedness\cite{Almeida2008,Bastolla2009,Allesina2013,Munoz2013} and assortativity\cite{Newman2002}), we have detected regular trends starting four years before 2007, which have been identified as early signals of a WTW structural change.

Our analysis shows that the global structure of trade has deeply changed during the years preceeding the 2007 worldwide crisis. In a nutshell, our results highlight that the bipartite WTW 1) starts becoming increasingly random, losing part of its internal structure, four years before the burst of the crisis in 2007 and 2) stabilises to a stationary level of randomness during the following three years. This suggests that 1) the truly crucial phase experienced by the WTW is represented by the quinquennium 2003-2007 (the year 2007 is just the ending point of such a phase started earlier in time) and that 2) during the period 2007-2010, the level of randomness of the WTW seems to have reached a stationary regime, leading us to conclude that a stabilization of the WTW structural change has been observed. Interestingly, this second result matches the comments reported in \cite{nber,nber2,nber3} on the first evidences of the crisis attenuation in the biennium 2009-2010, even if additional data are needed in order to understand if a trend inversion of the monitored quantities towards the recovery of the pre-crisis values is detectable from 2010 on.

Moreover, the analysis of the so-called macrosectors and of several countries-specific trade activity points out that very different tendencies co-exist within the WTW, evidencing that the most robust indicators of the WTW structural change are provided by the most volatile sectors (e.g. manufacturing and basic metals sectors) and by those countries which can be considered as emerging economies. Because of the ongoing industrialization process and the lack of a solid economy, the latter seem to be the most sensitive to the cycles of the worldwide economy.

Notably, the detection of a dramatic change in the network topology on a data set as the one considered for the present analysis, allows us to compare the behaviour of an economic system as the WTW with that of purely financial systems (as the interbank networks studied so far) in the years preceding the crisis. What emerges from our analysis is that economic and financial systems may show remarkably different reactions to critical events. For instance, the analysis of $z$-scores of the monopartite motifs has revealed that the Dutch Interbank Network (DIN) has undergone a structural change moving in the \emph{opposite}  direction to the one shown by our analysis here, i.e. departing from a phase which is compatible with the prediction of the monopartite configuration model two years before 2007 and showing progressively more significant signals of structural organization from then on \cite{myearly}. Unfortunately, the data set considered in \cite{myearly} makes it unfeasible to assess the post-2008 evolution of the DIN, even if the last data points describing the evolution of all dyadic - and some of the triadic - $z$-scores seem to indicate a trend inversion of the quantities of interest \cite{myearly}. Such an opposite reaction to the crisis of 2007 seems to confirm the differences between these two systems, already highlighted by similar analyses \cite{mystationary}.

Recently, a long debate has taken place in the economic literature about the effects of the 2007 financial crisis on the worldwide economy \cite{nber,nber2,nber3,food,food2,rein,french,lev}. Although our findings on the dynamics of the sectors considered for the present analysis seem to move in the same direction of the comments reported in \cite{nber,nber2,nber3,french,wood,food,food2} any possible claim about the mechanisms responsible of the contagion of the export market is beyond the aim of the present article. Nevertheless, in \cite{patt} it is pointed out that a strong overlap between countries with trade and financial sectors of systemic importance has indeed been observed: such countries are argued to be those with the highest potential for having transmitted disturbances to systemic stability, via either the trade or financial channels \cite{patt}. Testing such hypothesis and the relation between these two propagation channels, however, needs the analysis of a multilayer network, an important subject for further investigations.

\section*{Methods\label{methods}}

\subsection*{The Bipartite Configuration Model}

As in \cite{mybipmethod}, we define a grandcanonical ensemble of bipartite, undirected, binary networks in which the two layers have respectively $C$ and $P$ nodes. The degree sequence of nodes (i.e. the number of connections for each node, on both layers) is constrained to match, on average, the observed one. Using such a recipe, every graph $\mathbf{M}$ in the ensemble is assigned a probability coefficient reading 
\begin{equation*}
P(\mathbf{M}|\vec{x}, \vec{y})=\prod_cx_c^{d_c(\mathbf{M})}\prod_py_p^{u_p(\mathbf{M})}\prod_{c, p}(1+x_cy_p)^{-1},
\end{equation*}
depending on the countries degree (also known as \emph{diversification} \cite{Tacchella2012}) $d_c,\:c=1\color{magenta},\color{black}\dots\color{magenta},\color{black}C$ and the products degree (also known as \emph{ubiquity} \cite{Tacchella2012}) $u_p,\:p=1\color{magenta},\color{black}\dots\color{magenta},\color{black} P$, $x_c$ and $y_p$ being the Lagrange multipliers associated to the constrained degrees.

Constraining the ensemble average values of diversification and ubiquity induces the following link-specific probability (i.e. the probability that a link exists between country $c$ and product $p$ independently of the other links)

\begin{equation}
p_{cp}=\frac{x_cy_p}{1+x_cy_p}.
\label{prob}
\end{equation}

The numerical values of the unknown parameters $\vec{x}$ and $\vec{y}$ have to be determined by solving the following system of $C+P$ equations, which constrains the average values of countries diversification and products ubiquities to match the observed values,

\begin{eqnarray}
\langle d_c\rangle&=&d_c^*,\:c=1\dots C\nonumber\\
\langle u_p\rangle&=&u_p^*,\:p=1\dots P
\label{sys}
\end{eqnarray}

\noindent $\{d_c^*\}_{c=1}^C$ being the observed degree sequence of countries, $\{u_p^*\}_{p=1}^P$ being the observed degree sequence of products and $\langle \cdot\rangle$ representing the ensemble average of a given quantity, over the ensemble measure defined by eq. \ref{prob} - as $\langle d_c\rangle=\sum_pp_{cp}$ and $\langle u_p\rangle=\sum_cp_{cp}$. In what follows, the numerical values of the parameters satisfying eq. \ref{sys} will be indicated with an asterisk, ``$\ast$".

Once the ensemble distribution has been defined, our method can be implemented to calculate the expected value of the quantities of interest over the ensemble. In the present work we have sampled the grancanonical ensemble of binary, undirected, bipartite networks induced by the BiCM, according to the probability coefficients $P(\mathbf{M}|\vec{x}^*, \vec{y}^*)$ and calculated the average and variance of the generic quantity $X$ as sampling moments \cite{mysampling}

\begin{eqnarray}
\langle X\rangle&\simeq&\sum_{\mathbf{M}}\tilde{P}(\mathbf{M})X(\mathbf{M}),\\
\sigma^2_X&\simeq&\sum_{\mathbf{M}}\tilde{P}(\mathbf{M})[X(\mathbf{M})-\langle X\rangle]^2
\end{eqnarray}

\noindent according to the sampling frequencies $\tilde{P}(\mathbf{M})$ (the latter being the sampling-induced approximations of the ensemble frequencies $P(\mathbf{M})$.

The accuracy of the BiCM prediction in reproducing the value of quantity $X$, measured on the real network characterized by the observed biadjacency matrix $\mathbf{M}^*$, can be quantified by calculating the $z$-score, defined as

\begin{equation}
z_X=\frac{X(\mathbf{M}^*)-\langle X\rangle}{\sigma_X}.
\end{equation}

Whenever the ensemble distribution of the quantity $X$ closely follows a Gaussian, $z$-scores can be attributed the usual meaning of standardized variables, enclosing the $99.7\%$ of the probability distribution within the range $z_X\in[-3,3]$: any discrepancy between observations and expectations leading to values $|z_X|>3$ can thus be interpreted as statistically significant. Even if $z$-scores have been recognized to be dependent on the network size \cite{Milo2004}, our data set collects matrices with very similar volume, leading us to assume this effect to be negligible for making inference on the network evolution. 

On the other hand, whenever the ensemble distribution of the quantity $X$ deviates from a Gaussian (e.g. whenever we deal with rare events) $z$-scores cannot be interpreted in the aforementioned straight way and an alternative procedure to make statistical inference is needed: in these cases we have computed the box plots. Box plots are intended to sum up a whole probability distribution by showing no more than five percentiles: the 25th percentile, the 50th percentile and the 75th percentile (usually drawn as three lines delimitating a central box), plus the $0.15$th and the 99.85th percentiles (usually drawn as whiskers lying at the opposite sides of the box). Box plots can thus be used not only to provide a description of the ensemble distribution of the quantities of interest (in the present work, sampled by drawing 5000 matrices), but also to assess the statistical significance of their observed value against the null hypothesis represented by the BiCM. In fact, inference can be made once the amount of probability not included in the range spanned by the whiskers is known (in our case, at 99.7\% confidence level).

\subsection*{Bipartite motifs}

In the mathematical literature, our motifs are also known as \emph{bicliques}, i.e. complete subsets of bipartite graphs, where every node of the first (sub-)layer is connected to every node of the second (sub-)layer. In the jargon of graph theory, X-motifs are known as $K_{2,2}$ graphs, W-motifs are known as $K_{3,2}$ graphs and M-motifs are known as $K_{2,3}$ graphs (the first index referring to the countries layer and the second index referring to the products layer). Interestingly, they have also been shown to provide a meaningful insight into the organization of biological networks (as genetic, proteic, epidemic ones) \cite{bicliques}.

As stated in Results section, V- and $\Lambda$-motifs provide a rather limited information on the structure of the bipartite WTW: a proof of this statement follows. Let us recall the definition of V- and $\Lambda$-motifs: $N_V(\mathbf{M})=\sum_{c<c'}V_{cc'}$ with $V_{cc'}=\sum_pm_{cp}m_{c'p}$ and $N_\Lambda(\mathbf{M})=\sum_{p<p'}\Lambda_{pp'}$ with $\Lambda_{pp'}=\sum_cm_{cp}m_{cp'}$. The $z$-score of V- and $\Lambda$-motifs can be computed analytica; following \cite{mybipmethod} and considering that the bipartite WTW is quite sparse across the whole temporal interval (having, on average, link density $c=\frac{L}{C\cdot P}\simeq 0.13$, with a variation of $\sigma_c\simeq0.03$), the expressions for $z_V$ and $z_\Lambda$ simplify to $z_V\simeq \frac{P}{2\sqrt{L}}$ and $z_\Lambda\simeq \frac{C}{2\sqrt{L}}$, i.e. to functions of $L$, which shows a unique trend inversion in correspondence of the year 2007, whence their little informativeness about early structural changes.

\section*{Appendix}

\subsection*{The Revealed Comparative Advantage (RCA) threshold}

%\begin{figure}[b!]
%\begin{center}
%\includegraphics[width=0.48\textwidth]{Obs_selected_Lambda_legenda.pdf}
%\caption{List of sectors that we have analyzed in the present paper, according to the UN categorisation \cite{onu}.}
%\label{sectors}
%\end{center}
%\end{figure}

In order to retain only the relevant export data, we adopted the Revealed Comparative Advantage (RCA) threshold \cite{Tacchella2012}

\begin{equation}
\mbox{RCA}_{cp}\equiv\frac{\frac{q_{cp}}{\sum_{p'}q_{cp'}}}{\frac{\sum_{c'}q_{c'p}}{\sum_{c'p'}q_{c'p'}}}=\frac{q_{cp}}{d_c^w}\frac{W}{u_p^w},
\end{equation}

\noindent where $q_{cp}$ is the amout of product $p$ exported by country $c$ and the sums $\sum_{p'}q_{cp'}=d_c^w$ and $\sum_{c'}q_{c'p}=u_p^w$ indicate, respectively, the weighted diversification of country $c$ \cite{Tacchella2012} and the weighted ubiquity \cite{Tacchella2012} of product $p$. The recipe provided by RCA allows us to select the product $p$ whose impact on the export basket of country $c$ is larger than the impact of $p$ on the global market. If $\mbox{RCA}_{cp}\geq1$, $q_{cp}/d_c^w\geq u_p^w/W$ and the corresponding entry of the binary matrix $\mathbf{M}$ is $m_{cp}=1$; on the other hand, $\mbox{RCA}_{cp}<1$ implies that $m_{cp}=0$.

\subsection*{Additional sectors analysis}

\begin{figure*}[t!]
\begin{center}
\includegraphics[width=\textwidth]{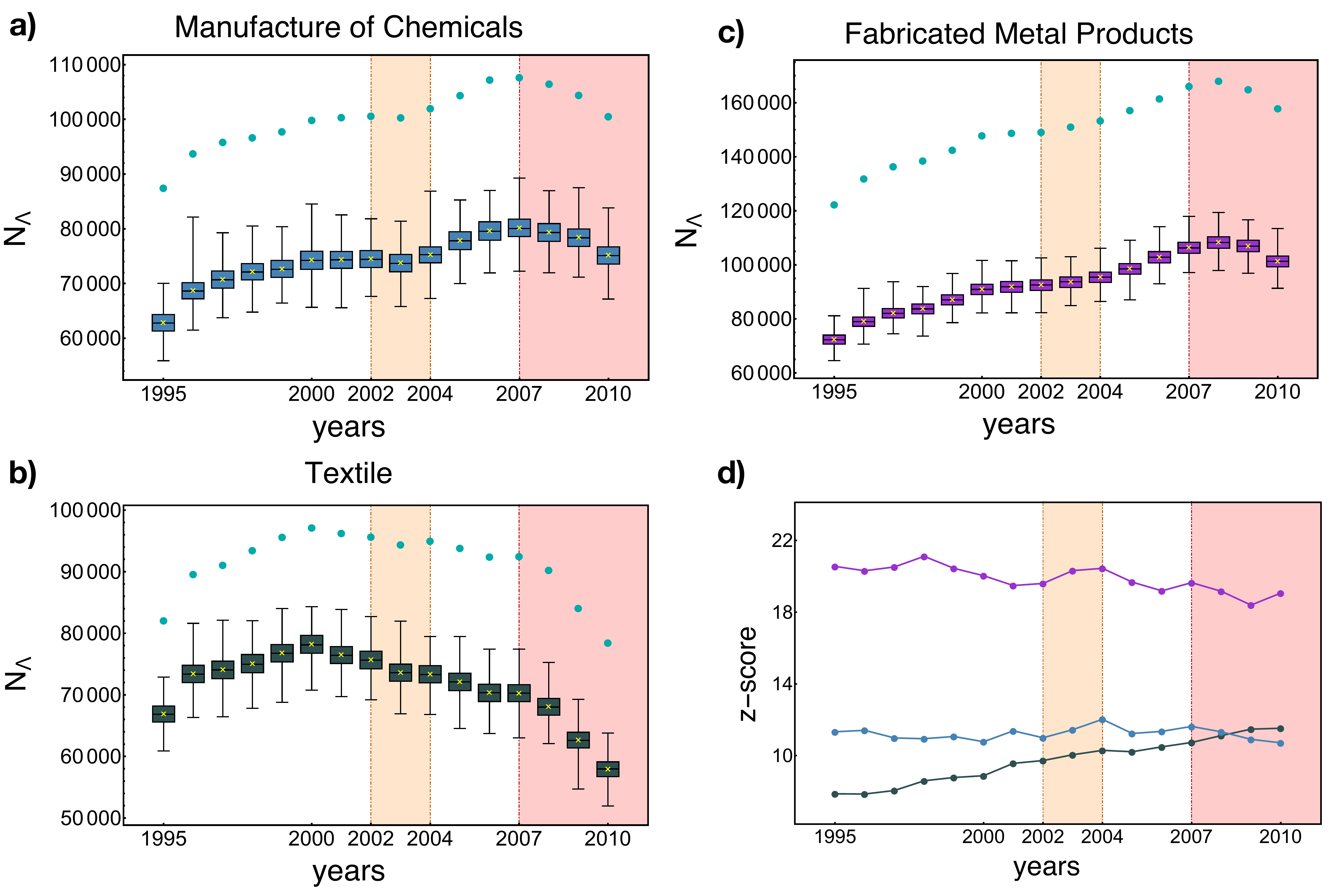}
\caption{Evolution of restricted $\Lambda$-motis across the period 1995-2010. Panels a), b) and c) show the box plots of the $\Lambda$-motifs of sectors \textcolor{Violet}{$\bullet$} - manufacture of chemicals, \textcolor{MidnightBlue}{$\bullet$} - manufacture of fabricated metal products and \textcolor{OliveGreen}{$\bullet$} - textile across the considered time window. The mean of the year-specific BiCM-induced ensemble distribution is indicated as a yellow cross; light-blue dots represent the observed abundances of motifs. Panel d) shows the evolution of the motifs $z$-scores (lines joining the dots simply provide a visual aid). The macrosectors characterized by the highest values of internal similarity seem to be insensitive to the approaching crisis, showing remarkably stable (increasing or decreasing) tendencies throughout our whole temporal window.}
\label{productsz3}
\end{center}
\end{figure*}

Although the macrosectors  ``manufacture of chemicals'', ``manufacture of fabricated metal products'' and ``textile'' are characterized by the highest values of internal similarity (thus matching better the definition of ``macrosectors'' in the economic jargon), they seem to be insensitive to the approaching crisis, showing remarkably stable tendencies throughout our whole temporal window (see fig. \ref{productsz3}). As shown in the section Results, the most informative sectors on the approaching crisis are those characterized by a moderately high level of internal similarity.

\subsection*{Additional measures for binary, undirected, bipartite networks}

In the following subsection, we provide the explicit definition of two additional topological quantities, in order to corroborate our findings with the analysis of more traditional network quantities. By applying the same kind of analysis presented in the main text, we show that early-warning signals of the crisis are detectable even analysing nestedness and assortativity.

\paragraph*{\underline{Nestedness.}} Loosely speaking, the degree of ``triangularity'' of the observed biadjacency matrix can be quantified according to a number of measures recently proposed under the common name of nestedness \cite{Almeida2008,Bastolla2009,Allesina2013,Munoz2013}. Here we adopt the one proposed in \cite{Almeida2008}, called NODF, an acronym for ``Nestedness metric based on Overlap and Decreasing Fill''.

Since the total value of nestedness ($\mbox{NODF}_t$) is the weighted average of the contribution from rows ($\mbox{NODF}_r$) and the contribution from columns ($\mbox{NODF}_c$), we have considered all of them for the present analysis. Naturally, the adopted measure of nestedness does not depend on the rows and columns ordering criterion \cite{Almeida2008}. If we indicate with $N_c$ the number of columns and with $N_r$ the number of rows, then

\begin{equation}
\mbox{NODF}_t\equiv\frac{N_c(N_c-1)\mbox{NODF}_c+N_r(N_r-1)\mbox{NODF}_r}{N_c(N_c-1)+N_r(N_r-1)}
\end{equation}

\noindent whose row- and column-specific contributions are provided by

\begin{equation}
\mbox{NODF}_c\equiv\frac{2\sum_{c<c'}T^C_{cc'}}{N_c(N_c-1)}; \:\:\mbox{NODF}_p\equiv\frac{2\sum_{p<p'}T^P_{pp'}}{N_p(N_p-1)},
\end{equation}

\noindent where $T^C_{cc'}=\frac{\sum_pm_{cp}m_{c'p}}{\mbox{min}[d_c,\:d_{c'}]}$ (if $d_c\neq d_c'$ and 0 otherwise) and $T^P_{pp'}=\frac{\sum_cm_{cp}m_{cp'}}{\mbox{min}[u_p,\:u_{p'}]}$ (if $u_p\neq u_p'$ and 0 otherwise) \cite{Almeida2008}.

\paragraph*{\underline{Assortativity coefficient.}} The second global quantity we have considered to characterize the WTW evolution is the assortativity measure $r\in[-1,1]$ proposed in \cite{Newman2002}, with $r=1$ indicating the maximum observable correlation between degrees (thus measuring the strongest tendency of links to connect nodes with similar degrees) and $r=-1$ indicating the minimum observable correlation between degrees (thus measuring the strongest tendency of links to connect nodes with different degrees).

The definition of the assortativity coefficient can be found in \cite{Newman2002}. By making it explicit which links contribute to the sums at the numerator and at the denominator, such a coefficient can be rewritten more clearly in terms of the bipartite nodes degrees, as

\begin{equation}
r\equiv\frac{4L\left(\sum_{c,p}m_{cp}d_cu_p\right)-\left(\sum_cd_c^2+\sum_pu_p^2\right)^2}{2L\left(\sum_{c}d_c^3+\sum_pu_p^3\right)-\left(\sum_cd_c^2+\sum_pu_p^2\right)^2}.
\end{equation}

\begin{figure}[t!]
\begin{center}
\includegraphics[width=0.5\textwidth]{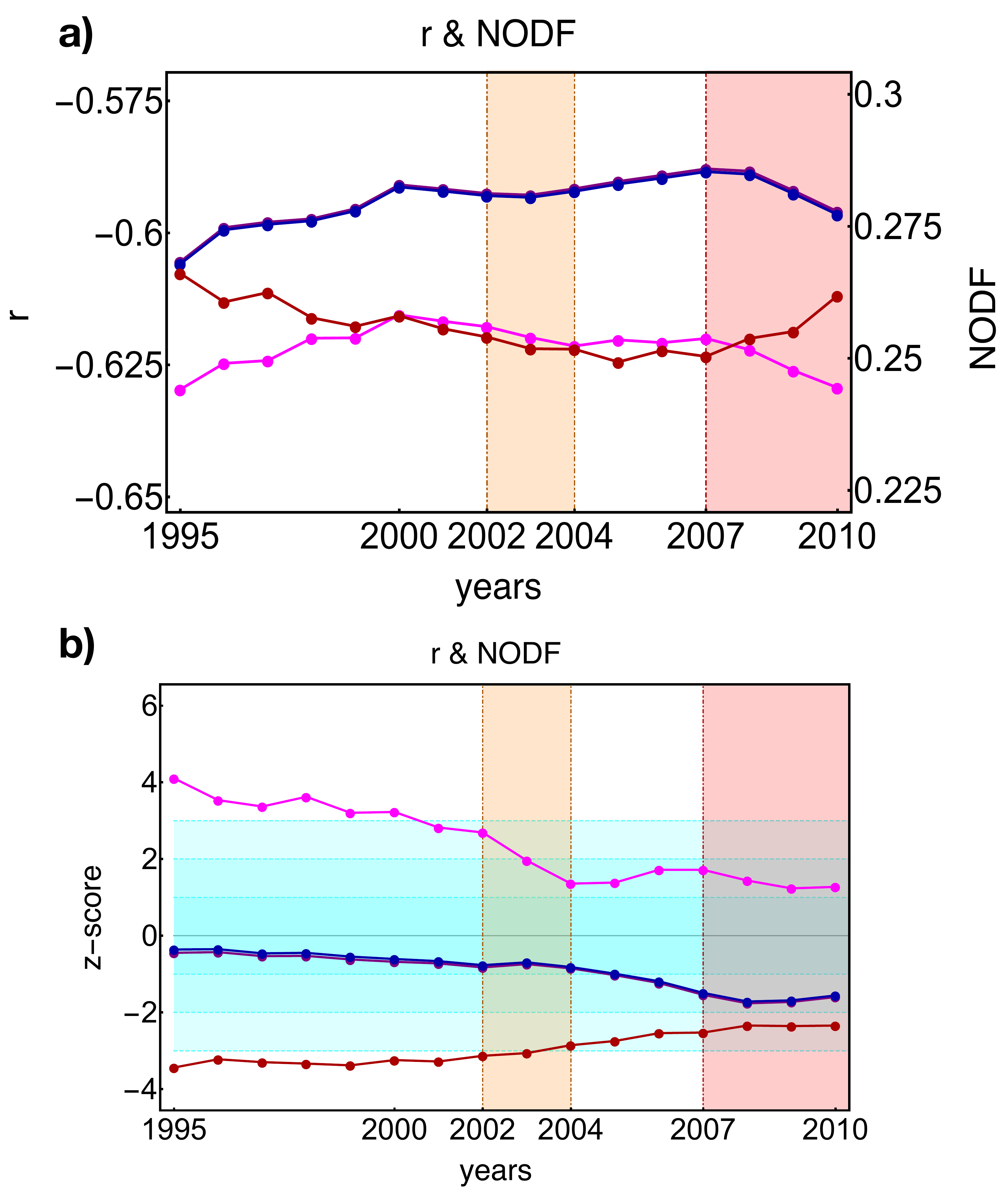}
\caption{Evolution across the period 1995-2010 of the total nestedness - \textcolor{blue}{$\bullet$}, the row-specific nestedness - \textcolor{Magenta}{$\bullet$}, the column-specific nestedness - \textcolor{Plum}{$\bullet$} and the assortativity - \textcolor{BrickRed}{$\bullet$}. Panel a) shows the evolution of the observed abundances; panel b) shows the evolution of the corresponding $z$-scores, pointing out the peculiarity of the year 2003 in discriminating between a statistically significant phase (1995-2003) and a phase consistent with our null model (2003-2007). Lines joining the dots simply provide a visual aid}.
\label{nodf}
\end{center}
\end{figure}

\paragraph*{\underline{Nestedness and Assortativity analysis.}} Fig. \ref{nodf}a shows the evolution of the observed abundances of nestedness (NODF$_t$ in blue; NODF$_r$ in pink; NODF$_c$ in purple) and assortativity (in brown). In particular, all of them provide evidence of the peculiar character of year 2007: all trends, in fact, even if characterized by opposite behaviors as the row-specific NODF and assortativity, show an inversion in correspondence of it.

Let us now carry on the comparison between the aforementioned observed trends and their expected counterparts under the BiCM, by plotting the corresponding $z$-scores. As shown by fig. \ref{nodf}b, NODF$_t$ and NODF$_c$ are always consistent with our null model, even if they gradually come closer to the $z=-2$ border as 2007 approaches. Again, the rate of increase of randomness across and after 2003 is notable. In particular, the total NODF spans one sigma of statistical significance in just four years (2003-2007), the same range having been spanned across the previous nine years.

As for their observed counterparts, the assortativity $z$-score is characterized by an opposite trend; in particular, it provides a clear statistical signal in 2003 by crossing the significance bound $z=-3$. This means that the degree of assortativity of the network becomes less and less significant, to become compatible with the BiCM-induced random value four years before 2007; it then steadily rises from 2003 until 2008 and seems to maintain such a value afterwards.

A more evident signal, confirming the increasingly random character of the network, is provided by the $z$-score of the row-specific NODF, whose analysis allows us to clearly distinguish two distinct phases, the first one lasting from 1995 to 2003 (characterized by a decreasing trend) and the second one lasting from 2003 to 2010 (characterized, instead, by an almost constant trend). The biennium across 2003 seems to constitute a somehow crucial period, defined by a decrease of statistical significance of two sigmas (from $z\simeq3$ to $z\simeq1$); analogously to what already observed for NODF$_t$, the same loss of statistical significance (from $z\simeq4$ to $z\simeq2$) was spanned by NODF$_r$ in the preceding eight years.

As for motifs, the $z$-scores trends considered so far agree in pointing out the peculiarity of 2003 as the year in which the network gets through two different regimes, from a ``structured'' phase, not compatible with our null model, to a increasingly random phase, where nodes correlations become increasingly similar to their random counterpart.

\begin{figure*}[t!]
\begin{center}
\includegraphics[width=0.99\textwidth]{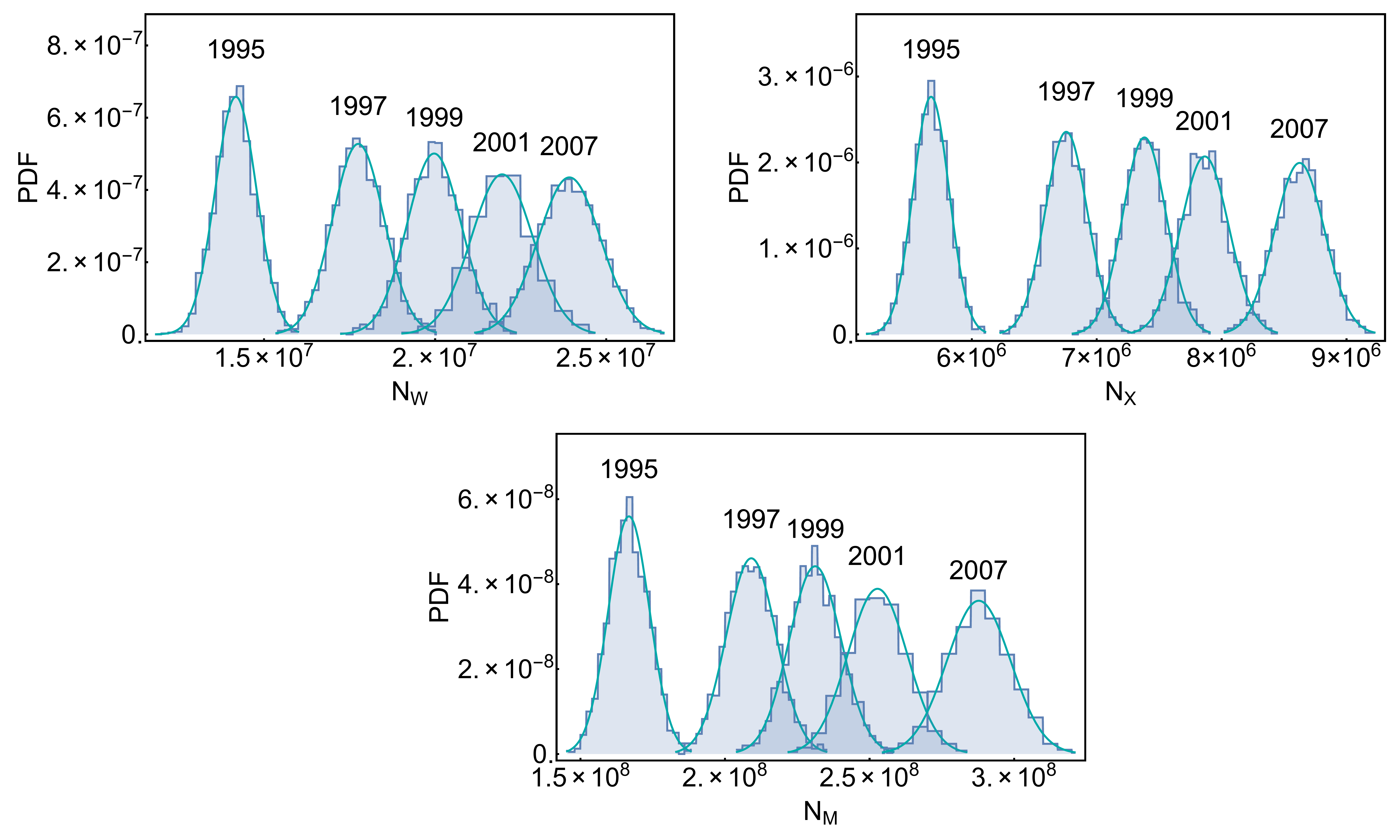}
\caption{Comparison of the ensemble distributions of our X-, W- and M-motifs with normal distributions having the same mean and variance, for different years of our data set. The very good agreement obtained justifies the computation of the $z$-score for the abundances of such motifs.}
\label{check}
\end{center}
\end{figure*}

\section*{Acknowledgements}

This work was supported by the EU project GROWTHCOM (611272), the Italian PNR project CRISIS-Lab and the EU project SIMPOL (grant num. 610704). The authors acknowledge Andrea Tacchella for the data cleaning and Giulio Cimini, Matthieu Cristelli, Emanuele Pugliese and Antonio Scala for useful discussions.

\section*{Author contributions statement}

FS and RDC analysed the data and prepared all figures. AG wrote the article. TS planned the research and wrote the article. All authors reviewed the manuscript.

\end{document}